\documentstyle[12pt,epsfig]{article}

\def\be{\begin{equation}}
\def\ee{\end{equation}}
\def\ba{\begin{array}{c}}
\def\ea{\end{array}}
\def\p{\partial}
\def\ben{$$}
\def\een{$$}
 
\begin{document}

\titlepage

  \begin{center}{\Large \bf
${\cal PT}$ symmetric models in more dimensions and solvable
square-well versions of their angular Schr\"{o}dinger equations
}\end{center}

\vspace{5mm}

  \begin{center}

Miloslav Znojil\footnote{ e-mail: znojil@ujf.cas.cz}

 \vspace{3mm}

\'{U}stav jadern\'e fyziky AV \v{C}R, 250 68 \v{R}e\v{z}, Czech
Republic\\

 \vspace{3mm}

\end{center}

\vspace{5mm}

\section*{Abstract}

From the partial differential Calogero's (three-body) and
Smorodinsky-Winternitz (superintegrable) Hamiltonians in two
variables we separate the respective angular Schr\"{o}dinger
equations and study the possibilities of their ``minimal" ${\cal PT}$
symmetric complexification. The simultaneous loss of the Hermiticity
and solvability of the respective angular potentials $V(\varphi)$ is
compensated by their replacement by solvable, purely imaginary and
piece-wise constant multiple wells $V_0(\varphi)$. We demonstrate
that the spectrum remains real and that it exhibits a rich ``four
series" structure in the double-well case.

\vspace{9mm}

\noindent PACS  03.65.Fd

\vspace{9mm}


 \newpage

\section{Introduction}

One of the most important methods of solution of partial differential
Schr\"{o}dinger equations in $D$ dimensions is an approximate
\cite{Zofka} or exact \cite{Fluegge} separation of variables in
spherical coordinates.  For example, once the potential is central, $
V(\vec{r}) =V(|\vec{r}|)$, the angular part of each wave function
coincides with the well known $D$ dimensional (hyper)spherical
harmonics.

In non-central case the explicit examples range from the confinement
of a single (quasi)particle in a superintegrable
Smorodinsky-Winternitz potential \cite{SmoWin} up to the binding of
$A>1$ particles in the exactly solvable models of the
Calogero-Moser-Sutherland type~\cite{Calogero,Sutherland}. Of course,
the structure of the corresponding $D$ dimensional generalizations of
the (hyper)spherical harmonics becomes more complicated. Our present
study was inspired precisely by this problem.

In section \ref{angSE} we shall introduce some basic notions and a
few explicit more or less standard examples of the above-mentioned
partial differential Schr\"{o}dinger equations (cf. subsection
\ref{angSE1} for more details). Then we shall specify our main
purpose, namely, the analysis of the models in question in their
appropriate non-Hermitian generalizations (a few remarks about
motivation will be collected in subsection \ref{angSE2}). In such a
setting, subsection \ref{angSE2} will summarize the key technical
aspects of our project.

In section \ref{toyM} we shall specify and explain in some detail the
present implementation of the Bender's and Boettcher's  idea
\cite{BB} of the weakening of the traditional Hermiticity to the mere
${\cal PT}$ symmetry. Subsection \ref{toyM1} will show that once we
decided to study just the two-dimensional partial differential
Schr\"{o}dinger equations, their angular part degenerates to the mere
single ordinary differential equation,
 \be
\left [ -\frac{d^2}{d\varphi^2} +V(\varphi) \right ] \,\psi(\varphi)
= \ell^2\,\psi(\varphi), \ \ \ \ \ \ \ \varphi \in (0,2\pi)\,.
\label{angular}
 \ee
This simplifies many technical aspects of our problem which may be
re-formulated as a confinement of a quasi-particle on a circle
\cite{QMoncircle}. Unfortunately, in contrast to the standard models
which use the Hermitian Hamiltonians and in which both the
interpretation and the solution of the angular Schr\"{o}dinger eq.
(\ref{angular}) remain straightforward, we encounter a really serious
difficulty here since the weakening of the Hermiticity  makes our
potentials $V(\varphi)$ unsolvable. As we shall show below, they all
acquire a multi-well shape with some $2M$ high barriers and deep
minima all over the circle.

Recently we ${\cal PT}$ symmetrized the three-body Calogero model
only via a very specific complexification of its angular part
(\ref{angular}) which enabled us to find new solutions and to write
them all down in closed form \cite{Tater}. This admitted a
preservation of many parallels between the Hermitian and
non-Hermitian case. Still, one cannot assert that our recipe was
fully satisfactory since, first of all, we did not provide any
guarantee that the construction of the solutions in polynomial form
does give us all the existing solutions. For this and other reasons
we are going to suggest now an alternative ${\cal PT}$ symmetrization
recipe.

The details will be described in subsection~\ref{toyM2} where we
postulate that all our angular potentials possess just a suitable
{\em piece-wise constant} shape which is, in addition, {\em purely
imaginary}. In this way, our choice will parallel many popular and
purely imaginary ${\cal PT}$ symmetric models  on the real line. {\em
Simultaneously}, we merely lose a small part of the simplicity of the
central models where the angular Schr\"{o}dinger equation may be
re-interpreted as a square well in quantum mechanics on a circle. For
a more detailed explanation of such a point of view, cf. the first
paragraph in subsection \ref{angSE1} below.

By construction, our multiple imaginary square wells on a circle
remain solvable by the matching method in principle. In practice,
this is illustrated in section \ref{solS} where we contemplate the
simplest nontrivial two-dimensional  problems of the above-mentioned
partial differential form. In technical terms, this section deals
with the angular Schr\"{o}dinger equation defined on a circle and is
a core of our present message. Its subsection \ref{solS1} describes
some details concerning the implementation of the current matching
method, and the subsequent subsection \ref{solS2} offers an ample
illustration of this recipe and its results. In three Tables we
sample the resulting spectra and discuss their apparent split in four
sub-series. Some methodical aspects of the practical search for these
energies are clarified with the assistance of several Figures. They
document the reliability of the method and show that just keeping in
mind a few simple rules concerning the (numerical) scaling, one may
calculate the energies with {\em any} pre-determined precision.

In our final section \ref{dis} we place all our results in a broader
context, i.a., of the popular double-well problems in subsection
\ref{dis1} and of their specific ``angular" features in subsection
\ref{dis2}.

\section{Angular Schr\"{o}dinger equations \label{angSE}}

\subsection{Hermitian non-central models \label{angSE1}}

Let us  recollect the simplest possible two-dimensional situation
where the angular Schr\"{o}dinger equation (\ref{angular}) is just
one-dimensional and defined on a circle (i.e., with the periodic
boundary conditions). In all the central cases it contains just the
trivial potential term, $V(\varphi) = 0$ for all $ \varphi \in
(0,2\pi)$. Hence, the well known (hyper)spherical solution of eq.
(\ref{angular}) is immediate, $\psi_m(\varphi) = \exp\, im\varphi$,
$\ell \equiv m = 0, 1, \ldots$.

In all the less trivial but still fully separable and exactly
solvable models with non-central potentials, the angular
Schr\"{o}dinger equation (\ref{angular}) remains exactly solvable by
assumption. One may recollect one of the oldest and simplest
illustrative examples of this type, viz., the Smorodinnsky-Winternitz
``superintegrable" oscillator in two dimensions \cite{SmoWin},
  \be
\left [ -\frac{\p^2}{\p X^2} -\frac{\p^2}{\p Y^2} + X^2 + Y^2
+\frac{G}{X^2}
+\frac{G}{Y^2}
 \right ]
\,\Phi(X,Y)={\cal E} \,\Phi(X,Y)\,.
 \label{parc} \label{Wint}
  \ee
After a routine elimination of the uninteresting motion of the
 centre of mass, also
 the
partial differential representation of the Calogero's popular
three-body bound-state model \cite{Calogero} has the very similar
form
   \ben
    \left [ -\frac{\p^2}{\p
X^2} -\frac{\p^2}{\p Y^2}+X^2+Y^2+ \right .
  \een
   \be
 \left .
  +\frac{g}{2\,X^{2}}+\frac{g}{2\,(X-\sqrt{3}Y)^{2}}
+\frac{g}{2\,(X+\sqrt{3}Y)^{2}}
 \right ]
\,\Phi(X,Y)={\cal E} \,\Phi(X,Y) \,. \label{Calo}
  \ee
Due to the presence of the repulsive barriers at $G > 0$ and $g > 0$,
the angular parts of both these equations cease to be defined on the
corresponding hypersphere (i.e., on the circle). Thus, once we
distinguish the above two models by their respective ``label numbers"
$M=M_{Winternitz} = 2$ and $M=M_{Calogero}=3$, we may summarize that

  \begin{itemize}

\item [(a)]
the domain of our innovated harmonics $\psi(\varphi)$ shrinks to a
segment $\varphi \in (0,\pi/M)$ of the full circle,

\item [(b)]
at both the ends of this interval, the Dirichlet boundary conditions
must be imposed, due to the presence of a strong singularity there,

\item [(c)]
both the models (with the elementary P\"{o}schl-Teller angular
potentials) still remain exactly solvable, as expected, in terms of
Jacobi polynomials~\cite{Flueggedva}.

\end{itemize}

 \noindent
All the three comments (a) -- (c) remain valid also for an
oversimplified ``toy" model
  \be
\left [ -\frac{\p^2}{\p X^2} -\frac{\p^2}{\p Y^2}
 + X^2 + Y^2
 +\frac{\cal G}{X^2}
 \right ]
\,\Phi(X,Y)={\cal E} \,\Phi(X,Y)\,.
 \label{parc1} \label{Wint1}
  \ee
Its ``index" $M=M_{toy}=1$ is ``minimal" and the model is popular as
a not entirely trivial methodical guide~\cite{Calogero,Tater}. It
will play the same guiding role in what follows.

\subsection{A motivation for transition to non-Hermitian
Hamiltonians  \label{angSE2}}

In the recent literature, an increasing attention is being paid to
non-Hermitian Hamiltonians with real spectra. Mostly they are
characterized by their parity-times-time-reversal symmetry, $H =
{\cal PT} H {\cal PT}$ \cite{BB,Flueggedva,BC} or, in a more formal
setting, by their specific parity-pseudo-Hermiticity property $H =
{\cal P} H^\dagger {\cal P}$ \cite{FV} -- \cite{Mostafazadeh}. Via
analytic continuation this tendency seems to offer a nontrivial
insight in the solutions~\cite{Taterdva}.

An important {\em phenomenological} appeal of the similar studies is
related to the observation that the complex conjugation operator
${\cal T}$ mimics the time reversal \cite{BBost,BBjmp}.  The
early stages of development of this subject were co-motivated by
its natural emergence within perturbation studies of various
models in quantum mechanics and field theory. People revealed,
for example, that even the peculiar and ``strongly
non-Hermitian" imaginary cubic anharmonic oscillator may have a
real and discrete spectrum bounded from below
\cite{Caliceti,Cal}. A few years later, an introduction of an
``unstable" quartic anharmonic oscillator proved {\em
mathematically} relevant in having facilitated the perturbative
description of some apparently non-perturbative double-well
models (e.g., of $H^{(DW)}(g) = p^2+x^2(1-g\,x^2)^2 + c\,x$ in
ref. \cite{BG}).  The weakening of the Hermiticity to mere
pseudo-Hermiticity was even found to make the non-Hermitian
quartic oscillators exactly solvable at certain couplings and
energies~\cite{BBjpa}.

The quick progress in the field has led to the sophisticated
perturbative analyses of some non-Hermitian ${\cal PT}$ symmetric
models in more dimensions~\cite{HH}. The ``analytic continuation"
activities intensify since the Hermitian non-central partial
differential Schr\"{o}dinger equations are attractive by the
variability of their possible {\em physical} interpretations.  They
might prove relevant in the models which range from a single particle
moving in a non-central $D$ dimensional potential \cite{Zhora} up to
a full-fledged $A$ body motion and spectra~\cite{Calogerodva}. Still,
the real impact of their non-Hermitian analytic continuations is not
yet fully predictable at present.

In contrast to the majority of contemporary publications which
study the ${\cal PT}$ symmetric models in one dimension, one
also has no clear guide to the purely technical aspects of the
${\cal PT}$ symmetrization of the models in more dimensions. We
feel motivated by this mathematical challenge and intend to pay
a deeper attention to the specific subset of the ${\cal PT}$
symmetrization techniques which is based on an ``incomplete"
complexification involving just the ``angular" part of the
problem.

{Without any detour to perturbation methods} we shall study some
specific properties of a class of the ``angular" Schr\"{o}dinger
equations in a way encouraged by our experience with the
comparatively easy tractability and insight provided by the
piece-wise constant potentials \cite{SQW}.

\subsection{Complexifications of the angular Schr\"{o}dinger
equations \label{angSE3}}

Within the framework of the so called ${\cal PT}$ symmetric quantum
mechanics \cite{BBjmp} one may complexify the model (\ref{Calo}) of
Calogero \cite{Tater,Basu} as well as its spiked and superintegrable
predecessors (\ref{Wint}) \cite{Jakubsky} by the complexification of
its single angular coordinate. Such an approach remains sufficiently
transparent. One of its very friendly features is that it
leads to a perceivable strengthening of the analogies between many
above-mentioned non-central models [with $g \neq 0$ in eq.
(\ref{Calo}) or with $G \neq 0$ in eq. (\ref{Wint})] and their
central, regular special cases (with $g=0$ or $G=0$, respectively).

In what follows, we just intend to study these analogies in more
detail. In a more explicit formulation of this purpose, our analysis
will start form the weakening of the Hermiticity to the mere ${\cal
PT}$ symmetry of the Hamiltonian. This opens a way towards a suitable
complex deformation of the integration path \cite{BB}, especially in
the vicinity of the singular points $\varphi_k=\pi\,k/M$, $k = 1, 2,
\ldots, 2M$. In this sense one should

  \begin{itemize}

\item [(A)]
re-extend the domain of the complexified harmonics $\psi(\varphi)$ to
the full circle,

\item [(B)]
return to the standard periodic boundary conditions in order to keep
our wave functions unambiguously defined,

\item [(C)]
re-analyze the solutions of eq. (\ref{angular}) in complex domain.

\end{itemize}

 \noindent
Unfortunately, none of the existing papers did solve all the three
problems (A) -- (C) at once. In particular, paper~\cite{paradoxes}
studied the solutions of eq. (\ref{angular}) on a single, isolated
complex segment, while a return to the full circle has been
facilitated in refs. \cite{Tater,Jakubsky} by the use of the
``non-periodic", Dirichlet boundary conditions.

An incompleteness of the latter results formed a key inspiration of
our present paper. In essence, it will circumvent some of the main
technical obstacles via a simple-minded replacement of the
complexified P\"{o}schl-Teller potentials in eq. (\ref{angular}) by
their square-well-like solvable approximants.

\section{Toy models \label{II} on a circle  \label{toyM}}

\subsection{A constant complex shift of the angle $\varphi$
\label{toyM1}}

Let us consider one of the above-listed angular Schr\"{o}dinger
equations (\ref{angular}) with the P\"{o}schl-Teller $M-$dependent
potential. One of its simplest complexifications on a circular domain
will be then determined by a mere constant complex shift (plus a
comfortable re-scaling) of the angular variable,
   \be
 \varphi=
 \varphi(s) = -i\,\alpha + \frac{\pi}{2}\,s,
 \ \ \ \ \ \ \alpha>0, \ \ \ s \in (-2,2)\,.
  \ee
Whenever the imaginary shift is very large, $\alpha \gg 1$, the
P\"{o}schl-Teller potential itself becomes very smooth and ``almost
forgets" about all its singularities along the real line,
   \be
 V^{(PT)}(\varphi) \sim 1/\sin^2 (M \varphi) =
 e^{-2M\alpha} \cdot
 \left [
 -\cos \pi M s + i\, \sin \pi M s + {\cal O} \left (
 e^{-2M\alpha}
 \right )
 \right ].
 \label{large}
  \ee
This potential is non-Hermitian and manifestly ${\cal PT}$ symmetric
(with the current definition of the complex conjugation ${\cal T}$
and parity ${\cal P} s = -s$). In generic case, its ``energies" [or
rather eigenvalues $\ell^2$ in eq. (\ref{angular})] may be expected
real~\cite{BBjmp}. In addition, its wave functions may be comfortably
expressed in terms of Bessel functions \cite{CJT}.

Unfortunately, the above solutions only become exact in the limit $\alpha
\to \infty$ where the potential (\ref{large}) itself vanishes.
Obviously, all the information about the original angular potential
is lost and nothing new is gained in the limit $\alpha \to \infty$
since, formally, one just returns to hyperspherical harmonics. {\it
Vice versa}, trying to preserve the specific,
$2M-$well character of our family of the
P\"{o}schl-Teller angular potentials $V^{(PT)}$ {\em we must return} to
the domain of the finite shifts $\alpha$. In what follows, we intend
to make just the first steps in this direction.

\subsection{Piece-wise constant angular potentials  \label{toyM2}}

Basically, we shall preserve the multi-well character of our
potentials, trying merely to deform the local details of their global
shape. In a more detailed specification of our ``toy-model" we were
guided by refs. \cite{periodicb} and \cite{periodicj}. In the former
one, the complex and ${\cal PT}-$symmetric potentials with elementary
trigonometric shapes resembling eq. (\ref{large}) have been analyzed
in {\em periodic} case. In accord with expectations, the
corresponding band spectrum proved real but in contrast to the
conventional real periodic potentials, it appeared to exhibit certain
significant {\em qualitatively new} features. Although, due to the
unsolvable character of the potentials, all these observations were
originally based just on the careful numerical analysis and
higher-order WKB techniques, their confirmation has been delivered in
the subsequent short note \cite{periodicj}. There, the presumably not
too relevant real part of the ``realistic" potential $V$ has been
replaced by a constant (zero). Moreover, the exact solvability of the
new schematic model has been achieved by an extremely drastic
reduction of all the remaining (i.e., purely imaginary) part of the
potential $V$ to the mere sequence of delta functions (known as the
Kronig-Penney model) with purely imaginary (i.e., still ${\cal
PT}-$symmetric) strengths mimicking the presence of the infinitely
many minima in the original $V$.

Here we shall proceed in a very similar manner. Firstly, we shall
ignore again the real part of the potential as not too relevant.
Besides all the above-mentioned indirect motivation of such a
technical simplification, we also directly verified its consistency
in ref. \cite{preceding} where a ``minimal" {\em non-periodic}
double-well version of the Kroning-Penney potential has been used.

Secondly, we shall keep in mind that the most important feature of
the remaining, purely imaginary part of our potentials (\ref{large})
lies in the presence of their {\em precisely} $2M$ separate extremes.
Thus, a transition to the solvable model will be achieved here via
their {\em piece-wise constant} approximation. In this manner, once
we denote $h = 1/M$, we arrive at the angular Schr\"{o}dinger eq.
(\ref{angular}) in the more or less unique form
   \be
\left [ -\frac{d^2}{ds^2} + V^{(M)}(s) \right ]\,\psi^{(M)}(s) =
E\,\psi^{(M)}(s)\,,
 \ \ \ \ \ \ \ \ \ s \in (-2,2)\,
 \label{angul}
  \ee
with the very specific class of the purely imaginary toy potentials
   \be
V^{(M)}(s) = \pm i\,Z,\ \ {\rm for}\ \  s \in  \triangle^{(\pm)}
 \label{toby}
  \ee
where
   \ben
 \triangle^{(+)}=
 (-2,-2+h)
 \bigcup (-2+2h,-2+3h) \bigcup \ldots \bigcup (2-2h,2-h),
 \een
and
   \ben
 \triangle^{(-)}=
 (-2+h,-2+2h)
 \bigcup \ldots \bigcup (2-3h,2-2h) \bigcup
  (2-h,2) \,.
    \een
For illustration, Figure {1} shows the shape of such a potential
defined over the circle at the Calogero's choice of~$M=3$.

In what follows we shall consider our potentials (\ref{toby}) and
concentrate on their first nontrivial special case with $M=1$,
   \be
 V(x) = \left \{
  \begin{array}{cc}
+i\,Z\ \ & x \in
 (-2,-1)
\bigcup (0,1),
\\
-i\,Z\ \ & x \in
 (-1,0)
\bigcup (1,2) \ea \right ..
 \label{equ}
 \label{toy}
  \ee
This means that a combination of the ${\cal PT}$ symmetry requirement
with the double-well character of the potential makes our $M=1$ model
defined on a circle of the convenient radius $R = 2/\pi$.

\section{Solutions \label{solS}}

\subsection{The sample of the matching
conditions at $M = 1$ \label{solS1}}

The explicit solutions of our toy example (\ref{toy}) will be
constructed by the matching method. For this purpose, it makes sense
to denote the left and right intervals $(-2,0)$ and $(0,2)$ by the
respective subscripts $_L$ and $_R$.  The subintervals which lie near
or far from the origin at $s=0$ will be assigned another subscript
$_N$ or $_F$, respectively. In this notation we have the natural ansatz
  \be
\psi_j(x) = A_j\sin \kappa_j x + B_j \cos \kappa_j x,
\ \ \ \ \ \ \ \ \ \ \ j = FL, NL, NR, FR
 \ee
with
   \be
\kappa_{FL}= \kappa_{NR}= \kappa=s-i\,t, \ \ \ \ \kappa_{NL}=
\kappa_{FR}= \kappa^*=s+i\,t, \ \ \ \ s > 0, \ \ \ t > 0.
  \ee
This should be interpreted as a mere suitable reparametrization of
the desired spectrum of ``energies"
   \be
E \equiv \ell^2= s^2-t^2, \ \ \ \ \ \ \ 2st \equiv Z > 0.
\label{fyzika}
  \ee
Now its is easy to require the matching of the logarithmic
derivatives at the discontinuity at $x=-1$,
   \be
\psi_{FL}(-1) = \psi_{NL}(-1), \ \ \ \ \ \ \ \ \
\psi'_{FL}(-1) = \psi'_{NL}(-1)\,,
  \ee
and, {\em mutatis mutandis}, at the other two ``internal" boundaries
at $x = 0$ and $x=1$. The points $x=2$ and $x=-2$ must be
matched together in a way which reflects the circular symmetry of the
problem,
   \be
\psi_{FR}(2) = \psi_{FL}(-2), \ \ \ \ \ \ \ \ \ \psi'_{FR}(2) =
\psi'_{FL}(-2)\,.
 \ee
We may conclude that the elementary insertions lead to the linear
system of eight homogeneous equations for all the eight coefficients
$A_j$ and $B_j$ (which may be arbitrarily normalized). This means
that the secular determinant of the system must vanish,
  \be
\det Q(s,t)=0\,.
\label{vanish}
 \ee
After one inserts, say, $s=s(t)= Z/2t$ from eq. (\ref{fyzika}), the
insertion of all the non-vanishing matrix elements leads to the
closed and compact form $F(t)=0$ of eq. (\ref{vanish}) with
  \be
Q_{11}=-Q^*_{13}=
-Q_{55}=
Q^*_{57}=
-\sin \kappa
 \ee
  \be
Q_{12}=-Q^*_{14}=
Q^*_{34}=-Q_{36}=
Q_{56}=
-Q^*_{58}=
\cos \kappa,
 \ee
  \be
Q_{21}=-Q^*_{23}=
Q^*_{43}=-Q_{45}=
Q_{65}=
-Q^*_{67}=
\kappa \cos \kappa
 \ee
  \be
Q_{22}=-Q^*_{24}=
-Q_{66}=
Q^*_{68}=
\kappa \sin \kappa,
 \ee
  \be
Q_{71}=Q^*_{77}=
-\sin 2 \kappa
, \ \ \ \ \ \ \ \
Q_{72}=-Q^*_{78}=
\cos 2 \kappa,
 \ee
  \be
Q_{81}=-Q^*_{87}= \kappa \cos 2 \kappa , \ \ \ \ \ \ \ \
Q_{82}=Q^*_{88}= \kappa \sin 2 \kappa
  \ee
and
   \be
 F(t) = -8\,t^4 \, {\rm sinh}^6 t\,\sin^6 \left (
 \frac{1}{2t}
 \right ) + \ldots
 -\frac{1}{2t^4}\, {\rm cosh}^6t\,\cos^6
 \left (
 \frac{1}{2t}
 \right )\,.
  \ee
We did not display it here in full detail because with all its 30
separate real trigonometric terms, this explicit formula is quite
lengthy and its detailed inspection seems hardly instructive at all.
For this reason, we are rather going to display here a few
characteristic graphs of the function $F(t)$ instead.

\subsection{The sample of a graphical determination
of the spectrum \label{solS2}}

One should start in a weak coupling regime. Choosing the very small
$Z=1/10$ we get the sequence $t_0 = 0.2219819562$,
$t_1=0.03467067057$, $\ldots$ of the real roots of $F(t)$.  Due to
the fact that the function $F(t)$ itself has the elementary form, the
numerical values of these roots may easily and quickly be evaluated
with an arbitrary precision.  These values then determine the
low-lying energies $E_0 = 0.00153255$, $E_1 = 2.078577$ etc.

An increase of the coupling to $Z=1$ shifts the lowest roots
($t_0=0.6564195696$ etc) and energies ($E_0=0.1493123386$ etc)
upwards.  No changes in the pattern are observed, and the more
detailed description of the $Z=1$ results is given here in the
three Tables \ref{table1} - \ref{table3}.

These sample results reveal and demonstrate a certain quadruple
periodicity of the energy levels.  This merely reflects the
simplicity and regularity of the secular determinant $F(t)$ itself.
An inspection of an overall shape of this function clarifies the
structure of the spectrum in a way which complements very well the
explicit Tables. Thus, we may see in Figure {2} that beyond its
rightmost or ``maximal" root $t_0 \approx 0.66$, the curve $F(t)$
remains negative and decreases towards $-\infty$ very quickly.

In a way, the ground-state root $t_0$ has a certain ``exceptional"
status in having no natural ``neighbor".  For all the rest of the
spectrum we observe that the roots may be organized as certain
doublets $(t_{2k-1}, t_{2k})$ numbered by $k = 1, 2, \ldots$.  With
an increase of the index $k$, these partners in the doublet lie
closer and closer to each other. This is illustrated in Figure {3}
where an onset of the existence of the doublets is clearly visible
near $t_1= 0.3934710177$ and where we also notice a rapid growth of
the maxima of $F(t)$ which makes the numerical search for its zeros
slightly more complicated.  In particular, it is necessary to keep
the necessary re-scalings of $F(t)$ during its graphical analysis
under control.

For this purpose we used an algorithm based on an interplay between
the plotting and root searching steps. For example, the doubts over
Figure {3} concerning the existence of the root near $t \approx 0.16$
are easily settled by the magnification of this subinterval in Figure
{4}. In this way we were able to detect and scale-out the
comparatively quick increase of the density of the roots $t_n>0$ with
the increase of their subscripts.

A new phenomenon emerged in the domain of $n \gg 1$, which may be
even noticed in the Figure {4} itself.  One detects a systematically
growing difference between the behaviour of the doublets $(t_{2k-1},
t_{2k})$ at their even and odd indices $k$. The phenomenon is even
better visible after we move to the energies themselves. Thus, a
glimpse in Table \ref{table1} verifies immediately that only at the
even $k = 2m$ the quasi-degeneracy of the pairs of roots $(t_{2k-1},
t_{2k})$ has a real impact.  Only in this case it implies also the
similar quasi-degeneracy effect for the related energies, $E_{4m-1}
\approx E_{4m}$, $m \gg 1$.

The comparatively bigger distance between $t_{2k-1}$ and $t_{2k}$ at
the odd $k = 2m+1$ makes the numerical determination of these pairs
less difficult.  {\it Vice versa}, the quick decrease of distance
between these roots at the even $k=2m$ may be perceived, especially
in combination with the growth of the maxima of $F(t)$ at the smaller
$t \to 0$, as the key obstacle encountered during the precise
numerical localization of the energies. A useful tool of a
disentanglement of these two obstacles has been found in the
introduction of an auxiliary function ${L}(t) = {\rm log} [ |F(t)|
]$, the practical {\em numerical} picture of which is sampled in
Figure~{5}.

Two of its properties have to be emphasized. Firstly, the
absolute-value-part of the definition of $L(t)$ guarantees that the
logarithm remains well defined {\em and} that the changes of sign of
the original functions $F(t)$ on a very small interval become,
empirically, ``magnified".  The emergence of every quasi-degenerate
pair of zeros will be reflected by a very well visible ``artificial
bump" in $L(t)$ in place of a hardly visible doublet of nodes in an
improperly rescaled graph of the unmodified function $F(t)$ (compare,
e.g., Figures {4} and {5}).

Secondly, the sequence of the maxima of $L(t)$ will not grow that
quickly with the decrease of $t\ll 1$.  Hence, the growing density of
the roots is better represented by the logarithm, in spite of its, in
principle, unlimited decrease near the zeros of $F(t)$. Empirically,
the latter effect proves negligible in numerical practice, in the way
illustrated very persuasively by Figure~{5}.

We may summarize that the main experience gained during the graphical
solution of our sample secular eq. (\ref{vanish}) lies in the
necessity of an appropriate selection of the interval in which the
separate roots $t_n$ are being sought.  Of course, the ``good" size
of these intervals will depend on the coupling $Z$ so that in
general, one should never forget to consult the graphs of $F(t)$ {\em
and} $L(t)$ .

\section{Discussion \label{dis}}

\subsection{Multiple-well problems
 in non-Hermitian framework \label{dis1}}

We may summarize that a family of certain purely imaginary potentials
was considered on a circle. Their piece-wise constant shape was
chosen to mimic the multi-well angular part of certain not yet fully
understood pseudo-Hermitian versions of Calogero-Moser and
Smorodinsky-Winternitz Hamiltonians with real energies. Vice versa,
the reality of our spectrum reflects a ``minimal" square-well-type
${\cal PT}$ symmetrization of the initial many-body or
more-dimensional separable Hamiltonian. A thorough numerical analysis
of the spectrum was performed for the first nontrivial special case
with double well potential. It has been shown to exhibit a rich
``four series" structure.

These results fit well a broader perspective. Forty years ago,
perturbation analysis of the single-well oscillator $H^{(SW)}(g) =
p^2+x^2+g^2\,x^4$ revealed that the most natural interpretation of
its ``physical" spectrum $\{ \varepsilon_n(g)\}$ emerges via a
complexification of its coupling $g$ \cite{BW}.
{\em All} the apparently uncorrelated numbers $\varepsilon_n(g)$ were
shown to coincide with the values of a {\em single} analytic function
$ \varepsilon(g)$, evaluated on the {\em different} Riemannian
sheets. In this interpretation, the main quantum number $n = 0,
1,\ldots$ merely selects the different Riemannian sheets in a
certain specific, very transparent and intuitively appealing
manner. One may conclude that the complexification of the
parameters is one of the best ways towards an understanding of
the spectra, especially when they exhibit a strongly anharmonic
double well shape \cite{Seznec} since, in these cases, there may
exist a close connection between the single and double wells
mediated by Fourier transformation and analytic
continuation~\cite{BG}.

{\it A priori} one might expect that the high degree of an internal
symmetry of the superintegrable equations (which implies their
separability and exact solvability in several systems of coordinates
\cite{Zhora}) might mediate the facilitated technical feasibility
and, hence, preferrence of their systematic ${\cal PT}$
symmetrization. For this reason, our present recipe
might be re-read as a direct ``brute-force" introduction of a
Hermiticity-breaking term. Such a modification of the superintegrable
Schr\"{o}dinger equations by an additional potential might offer a
new insight in the problem of the reality of the spectra and, in
particular, in the (sub)problem of the mechanisms of its
possible (and, in principle, admissible and often quite easy)
breakdown.

\subsection{Summary and outlook  \label{dis2}}

The relationship of the ${\cal PT}$ symmetry of the Hamiltonian
to the reality of its spectrum was illustrated here via a few
most elementary non-Hermitian solvable models in two dimensions.
In the similar examples, the choice of the suitable ${\cal PT}$
symmetric anharmonic terms is usually dictated by the purely
technical requirements of simplicity and solvability. In this
sense, one encounters serious difficulties when moving to two
and more dimensions since only the one dimensional
Schr\"{o}dinger equations abound with the solvable potentials.
To a successful ${\cal PT}$ symmetric generalization of these
models to more dimensions, the key contribution of our present
short note may be seen in its proposal and verification of a
combination of the  matching and analytic techniques. We showed
that in spite of the violation of the superintegrability caused
by our introduction of the step-like imaginary potentials, we
were still able to study our models by the essentially
non-numerical technique.

The angular potentials we considered were drastically simplified
to a piece-wise constant purely imaginary field which manifestly
destroys the Hermiticity of the Hamiltonian. Our extreme
simplification of the generic situation offered the picture
where the barriers become penetrable. This changes the boundary
conditions to periodic ones \cite{Taterdva}. The related
analysis proved extremely illuminating and revealed the
existence of the new and hardly expected types of a regularity
in the spectrum. We may conclude that our choice of the
schematic toy model (\ref{toy}) was fortunate, indeed.

  \begin{itemize}

\item
Qualitatively, the energies in the spectrum seem to be classified as
lying in the four separate series.

\item
The intrinsic double-well character of the model seems to be
responsible for the emergence of the almost degenerate energy
doublets.

\item
One encounters the quasi-degeneracy of the energies not only in the
low energy domain but, quite unexpectedly, all over the energy scale.

\end{itemize}

 \noindent
There exist several open questions which remained unanswered.
Firstly, a deeper understanding of many empirical observations is
missing. The quadruple-series spectrum in Table~\ref{table1} as
re-emphasized in Tables~\ref{table2} and \ref{table3} might motivate
a continuing analysis of the more-dimensional systems.

Secondly, there is no doubt that the structure of the spectrum in our
present ${\cal PT}$ symmetric model is richer than in its Hermitian
predecessors. Although all the intuitive and/or elementary
interpretations of the ``angular-type" motion over the circle remain
similar, one feels tempted to emphasize the differences.  In this
sense, one notices, first of all, that our present model re-confirms
the idea and/or existence of certain ``quantum beats" as conjectured
on the basis of a different model in ref.~\cite{preceding}.

Thirdly, one has to re-emphasize that in contrast to the
quasi-degeneracy as observed in the Hermitian multiple wells, the
present, ${\cal PT}$-symmetric analogue of this phenomenon is
``periodic" and transcedes the standard limitation of the
quasi-degeneracy to the mere low-lying spectrum.

Finally, we might re-open the discussion which has been initiated in
ref. \cite{Taterdva}. An ambiguity of a return to the Hermitian case
has been revealed there, and interpreted as a transformation which
gives rise to the unexpected emergence of some entirely new models.
Of course, once we started here from a separation of a circle in some
$2M$ quasi-independent sectors, similar outcome might be expected in
principle. Indeed, our example just deals with the parts (quadrants)
of a bigger phase space. As long as this space was originally
decomposed in a few invariant subspaces, its non-Hermitization might
offer a recipe how one could open a tunnel through the originally
fully impenetrable barriers.

\section*{Acknowledgment}

The work was supported by GA AS in Prague, contract No. A 1048302.

\newpage

\section*{Table captions}

.

Table 1.
The low-lying energy spectrum at $Z = 1$.

Table 2.
Two natural subsets of the even energy
levels of Table \ref{table1}.

Table 3.
Two natural subsets of the odd energy
 levels of Table \ref{table1}.

\section*{Figure captions}

.

Figure 1.
Imaginary part of a Calogerian six-well
angular potential in its rectangular approximation.

Figure 2.
Monotonous decrease of the secular determinant $F(t)$ beyond
its ground-state root $t_0 \approx 0.656$ at $Z=1$.

Figure 3.
Graphical search for the roots of $F(t)$.

Figure 4. A magnification of Figure {3} near $t=0.16$.

Figure 5.
Logarithmic re-scaling ${L}(t) = {\rm log} [ |F(t)| ]$ and an
improved graphical trapping of quasi-degeneracy.

\newpage

   \begin{table}
\thispagestyle{empty}
 \caption{The low-lying energy spectrum at $Z = 1$}
\label{table1}
  \begin{center}
  \begin{tabular}{||c||ccc||}
  \hline\hline
 level     & parameter & energy & difference $\triangle_n$ \\
 $n$     & $t_n$ & $E_n$ & $ E_n-E_{n-1}  $  \\
\hline
\hline
0& 0.656
& 0.149312
& -
\\
1& 0.393
& 1.459965
&1.319
\\
2& 0.266
& 3.464686
&2.004
\\
3& 0.1596
& 9.792771
&6.328
\\
4& 0.1587
& 9.895111
&0.102
\\
5& 0.111
& 20.127356
&10.232
\\
6& 0.101
& 24.273237
&4.146
\\
7& 0.07959
& 39.459389
&15.186
\\
8& 0.07956
& 39.484770
&0.025
\\
9& 0.0651
& 58.954565
&19.469
\\
10& 0.0623
& 64.410749
&5.456
\\
11& 0.053053
& 88.817991
&24.397
\\
12& 0.053050
& 88.829258
&0.011
\\
13& 0.0461
& 117.64814
&28.819
\\
14& 0.0449
& 124.15474
&6.507
\\
15& 0.039789
& 157.90892
&33.754
\\
16& 0.039788
& 157.91526
&0.006
\\
17& 0.0357
& 196.15402
&38.239
\\
  \hline
\hline
\end{tabular}
\end{center}
\end{table}

\newpage

   \begin{table}
 \caption{Two natural subsets of the even energy
 levels of Table \ref{table1}}
\label{table2}
  \begin{center}
  \begin{tabular}{||c|c||c|c||}
  \hline\hline
 level     & second difference &
 level     & second difference \\ n
& $\triangle_n-\triangle_{n-4}$ & n
& $\triangle_n-\triangle_{n-4}$ \\
  \hline
\hline 6&2.142& 8&-0.77
\\
10& 1.310&
12& -0.014
\\
14& 1.051&
16& -0.0049
\\
  \hline
\hline
\end{tabular}
\end{center}
\end{table}

\newpage
   \begin{table}
 \caption{Two natural subsets of the odd energy
 levels of Table \ref{table1}}
\label{table3}
  \begin{center}
  \begin{tabular}{||c||c|c||}
  \hline\hline
 level     & second difference &
third difference \\ n
& $\triangle_n-\triangle_{n-2}$
& $\triangle_n-2\triangle_{n-4}+\triangle_{n-8}$ \\
  \hline
\hline
3& 5.018 &  -\\
5& 3.904&
$-$1.114
\\
7& 4.954&
+1.050
\\
9& 4.283&
$-$0.671
\\
11&4.938&
+0.655
\\
13& 4.422&
$-$0.516
\\
15& 4.935&
+0.513
\\
17& 4.485&
$-$0.450
\\
  \hline
\hline
\end{tabular}
\end{center}
\end{table}

\end{document}